\documentclass[aps,prstab,twocolumn,superscriptaddress]{revtex4}
\usepackage{graphicx}  % needed for figures
\usepackage{amsfonts}
\usepackage{amssymb}
\usepackage{amsmath}
\usepackage[colorlinks, linkcolor = black, citecolor = black, filecolor = black, urlcolor = black]{hyperref}  % \usepackage{hyperref} colored external/internal links
\usepackage{multirow}
\usepackage{newtxmath}
\usepackage{xfrac}
\usepackage{color}
\usepackage{makecell}
\usepackage{tabu}
\usepackage{upgreek}
\newcommand{\gsi}{\affiliation{GSI Helmholtzzentrum f\"ur Schwerionenforschung GmbH, Darmstadt, Germany}}
\newcommand{\guf}{\affiliation{Goethe-Universit\"at, Frankfurt am Main, Germany}}
\newcommand{\imp}{\affiliation{Institute of Modern Physics, Lanzhou, China}}
\newcommand{\edin}{\affiliation{University of Edinburgh, Edinburgh, United Kingdom}}
\newcommand{\atomki}{\affiliation{Institute for Nuclear Research (Atomki), Debrecen, Hungary}}
\newcommand{\cenbg}{\affiliation{CENBG, CNRS-IN2P3, Gradignan, France}}
\newcommand{\jlu}{\affiliation{Justus-Liebig Universit\"at, Gie{\ss}en, Germany}}
\newcommand{\anu}{\affiliation{Australian National University, Canberra, Australia}}
\newcommand{\hij}{\affiliation{Helmholtz-Institut Jena, Jena, Germany}}
\newcommand{\spbu}{\affiliation{St. Petersburg State University, St. Petersburg, Russia}}
\newcommand{\bau}{\affiliation{Al-Balqa' Applied University, Salt, Jordan}}

\newcommand{\tud}{\affiliation{Technische Universit\"at Darmstadt, Darmstadt, Germany}}
\newcommand{\mpi}{\affiliation{Max-Planck-Institut f\"ur Kernphysik (MPIK), Heidelberg, Germany}}

\newcommand{\ioq}{\affiliation{Institut f{\"u}r Optik und Quantenelektronik, Friedrich-Schiller-Universit{\"a}t, Jena, Germany}}

\begin{document}

\title{Electron capture of Xe$^{54+}$ in collisions with H${_2}$ molecules in the energy range between 5.5 MeV/u and 30.9 MeV/u}

\author{F. M. Kr{\"o}ger}
\email{felix.kroeger@uni-jena.de}
\hij
\gsi
\ioq
\author{G. Weber}
\hij
\gsi
\author{M.~O.~Herdrich}
\hij
\gsi
\ioq
\author{\mbox{J. Glorius}}
\gsi
\author{\mbox{C. Langer}}
\guf
\author{\mbox{Z. Slavkovsk\'a}}
\guf
\author{\mbox{L. Bott}}
\guf
\author{\mbox{C. Brandau}}
\gsi\jlu
\author{\mbox{B. Br\"uckner}}
\guf
\author{\mbox{K. Blaum}}
\mpi
\author{\mbox{X. Chen}}
\imp
\author{\mbox{S. Dababneh}}
\bau
\author{\mbox{T. Davinson}}
\edin
\author{\mbox{P. Erbacher}}
\guf
\author{\mbox{S. Fiebiger}}
\guf
\author{\mbox{T. Ga{\ss}ner}}
\gsi
\author{\mbox{K. G\"obel}}
\guf
\author{\mbox{M. Groothuis}}
\guf
\author{\mbox{A. Gumberidze}}
\gsi
\author{\mbox{Gy. Gy\"urky}}
\atomki
\author{\mbox{S. Hagmann}}
\gsi
\guf
\author{\mbox{C. Hahn}}
\hij
\gsi
\ioq
\author{\mbox{M. Heil}}
\gsi
\author{\mbox{R. Hess}}
\gsi
\author{\mbox{R. Hensch}}
\guf
\author{\mbox{P. Hillmann}}
\guf
\author{\mbox{P.-M. Hillenbrand}}
\gsi
\author{\mbox{O. Hinrichs}}
\guf
\author{\mbox{B. Jurado}}
\cenbg
\author{\mbox{T. Kausch}}
\guf
\author{\mbox{A. Khodaparast}}
\gsi\guf
\author{\mbox{T. Kisselbach}}
\guf
\author{\mbox{N. Klapper}}
\guf
\author{\mbox{C. Kozhuharov}}
\gsi
\author{\mbox{D. Kurtulgil}}
\guf
\author{\mbox{G. Lane}}
\anu
\author{\mbox{C. Lederer-Woods}}
\edin
\author{\mbox{M. Lestinsky}}
\gsi
\author{\mbox{S. Litvinov}}
\gsi
\author{\mbox{Yu. A. Litvinov}}
\gsi
\author{\mbox{B. L\"oher}}
\gsi
\tud
\author{\mbox{F. Nolden}}
\gsi
\author{\mbox{N. Petridis}}
\gsi
\author{\mbox{U. Popp}}
\gsi
\author{\mbox{M. Reed}}
\anu
\author{\mbox{R. Reifarth}}
\guf
\author{\mbox{M. S. Sanjari}}
\gsi
\author{\mbox{H. Simon}}
\gsi
\author{\mbox{U. Spillmann}}
\gsi
\author{\mbox{M. Steck}}
\gsi
\author{\mbox{J. Stumm}}
\guf
\author{\mbox{T. Sz\"ucs}}
\atomki
\affiliation{Helmholtz-Zentrum Dresden-Rossendorf (HZDR), Dresden, Germany}
\author{\mbox{T. T. Nguyen}}
\guf
\author{\mbox{A. Taremi Zadeh}}
\guf
\author{\mbox{B. Thomas}}
\guf
\author{\mbox{S. Yu. Torilov}}
\spbu
\author{\mbox{H. T\"ornqvist}}
\gsi\tud
\author{\mbox{C. Trageser}}
\gsi\jlu
\author{\mbox{S. Trotsenko}}
\gsi
\author{\mbox{M. Volknandt}}
\guf
\author{\mbox{M. Weigand}}
\guf
\author{\mbox{C. Wolf}}
\guf
\author{\mbox{P. J. Woods}}
\edin
\author{V. P. Shevelko}
\affiliation{P.N. Lebedev Physical Institute, Moscow, Russia}
\author{I. Yu. Tolstikhina}
\affiliation{P.N. Lebedev Physical Institute, Moscow, Russia}
\author{Th.~St{\"o}hlker}
\hij
\gsi
\ioq

\date{\today}

\begin{abstract}
The electron capture process was studied for Xe$^{54+}$ colliding with H$_2$ molecules at the internal gas target of the ESR storage ring at GSI, Darmstadt. Cross section values for electron capture into excited projectile states were deduced from the observed emission cross section of Lyman radiation, being emitted by the hydrogen-like ions subsequent to the capture of a target electron. The ion beam energy range was varied between 5.5 MeV/u and 30.9 MeV/u by applying the deceleration mode of the ESR. Thus, electron capture data was recorded at the intermediate and in particular the low collision energy regime, well below the beam energy necessary to produce bare xenon ions. The obtained data is found to be in reasonable qualitative agreement with theoretical approaches, while a commonly applied empirical formula significantly overestimates the experimental findings.

\end{abstract}

\pacs{34.50.Bw}

\keywords{Charge exchange cross sections, non-radiative electron capture, eikonal approximation, Schlachter formula}

\maketitle

\section{Introduction}
Charge-changing processes, i.e.\ loss or capture of electrons, occurring in ion-atom and ion-ion collisions belong to the most basic interactions for ion beams. Besides basic research, i.e. in atomic and plasma physics as well as astrophysics, the investigation of these processes is motivated by their paramount importance for many applications, such as ion stripping as well as transport and storage of ion beams in accelerator facilities~\cite{Stoehlker98,Shevelko12,Imao12}. In dispersive ion optical elements the trajectories of ions being up- or down-charged deviate from the one of the reference charge state, resulting in successive defocusing and eventual partial loss of the ion beam. Therefore, exact knowledge of the charge-changing cross sections in collisions with residual gas constituents or dedicated targets is of crucial importance for the planning of experiments at existing accelerators and storage rings as well as for the design of new facilities or upgrade programs.

This is particularly evident for the new Facility for Antiproton and Ion Research (FAIR)~\cite{Fortov2012}, currently under construction on the campus of the GSI Helmholtzzentrum f\"ur Schwerionenforschung in Darmstadt, Germany. At FAIR, future studies with heavy, highly charged ions will cover a previously unexplored range of experimental parameters with respect to collision energies, beam intensities and ion species. With the low-energy storage ring CRYRING@ESR (while being operated at Manne Siegbahn Laboratory of Stockholm University until 2009 this ring was referred to as CRYRING) the first experimental facility at FAIR was recently commissioned~\cite{Lestinsky2015}. This storage ring allows the storage of heavy, highly charged ions at beam energies between a few 100\,keV/u and roughly 10\,MeV/u. An important part of the research program of the Stored Particles Atomic Physics Research Collaboration (SPARC)~\cite{Stohlker2014} within the FAIR project concerns studies of highly charged ion beams that are decelerated in the CRYRING@ESR to energies significantly lower than previously accessible~\cite{Lestinsky2016} at the existing Experimental Storage Ring (ESR). These ions will have charge states much higher than their respective equilibrium charge state, and as a consequence electron capture from the residual gas constituents in CRYRING@ESR will be the dominant beam-loss process. Modeling these beam losses is of key importance for the efficient planning of experiments at CRYRING@ESR. However, experimental electron-capture cross-section data for highly-charged ions colliding with atoms/molecules at kinetic energies well below the respective projectile's ionization threshold is scarce. Therefore, additional experimental data covering a significant range of collision energies as well as ion species and target atoms is needed to benchmark theoretical approaches and scaling laws available for such collision systems. 

In this work, cross section measurements are presented for electron capture into excited states of initially bare xenon (Xe$^{54+}$) projectiles, occurring in collisions with hydrogen molecules~(H$_{2}$). This study was conducted using decelerated ion beams in the ESR, thus covering ion beam energies that are relevant for the operation of the CRYRING@ESR. The obtained electron-capture cross-section data is compared to combined predictions of the radiative electron capture (REC) and of the non-radiative electron capture (NRC) theories as well as to the widely used empirical Schlachter formula~\cite{Schlachter83}.

\section{Measurement technique and data analysis}
\begin{figure}[h]
\begin{center}
\includegraphics[width=8.6cm]{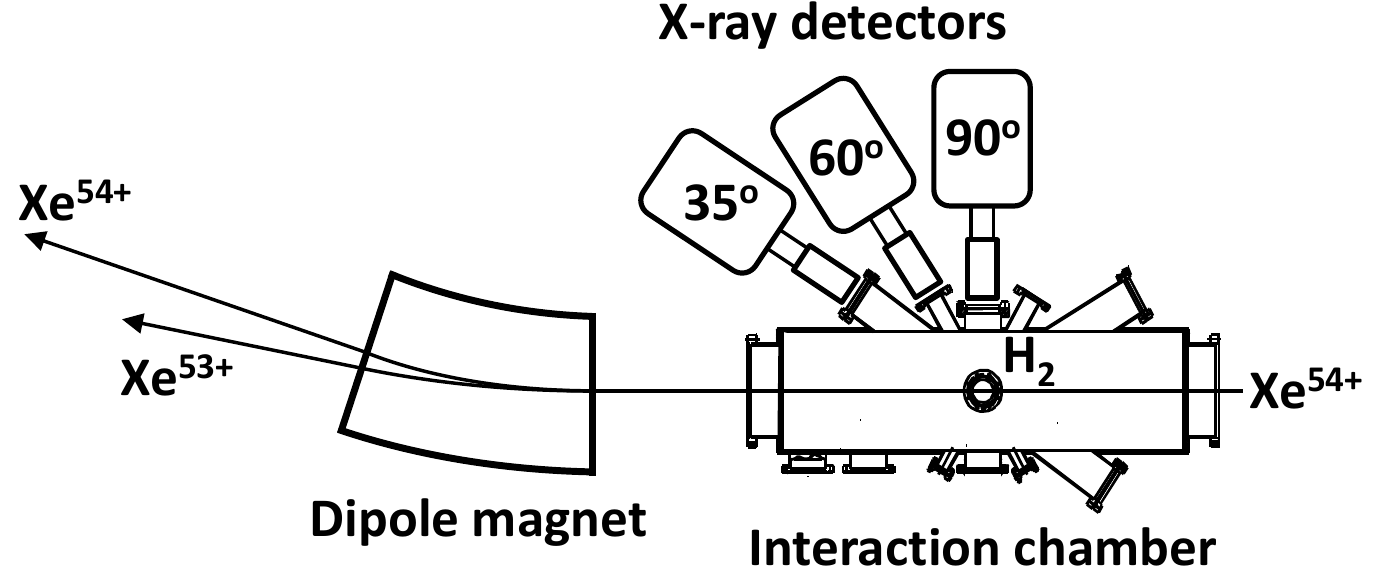}
\end{center}
\caption{Schematic drawing of the experimental setup. An array of HPGe X-ray detectors is positioned around the interaction zone of the ion beam and the gas jet target.
\label{fig1}}
\end{figure}

During a beam time at the GSI accelerator facility, xenon ions produced in an ion source at low charge states were pre-accelerated in the Universal Linear Accelerator (UNILAC) and subsequently injected into the heavy ion synchrotron SIS18. After being accelerated to a respective beam energy of 100\,MeV/u, the ions were ejected into the transfer line from SIS18 to the Experimental Storage Ring (ESR)~\cite{FRANZKE1987} where all their electrons were removed during the passage through an 11\,mg/cm$^2$ carbon stripper foil. After injection into the ESR, the ion beam was rebunched and decelerated, to the chosen final energies of 5.5, 6, 6.7, 7, 8, 15, and 30.9\,MeV/u. This energy range corresponds to relativistic $\beta$ values from 0.11 to 0.25. Typical beam intensities of several $10^7$~particles were stored and electron cooled in order to achieve a good beam quality, resulting in a typical beam diameter in the order of 2\,mm as well as a momentum spread in the order of $\Delta p / p \approx 10^{-4}$~\cite{Steck04} after a cooling time of a few seconds. As a next step the internal gas target of the ESR \cite{Grisenti2006,Kuehnel2009} was turned on leading to the formation of a H$_2$ gas jet with a diameter of about 6\,mm~\cite{Gassner15}. The overlap between the ion beam and the target was optimized based on a scan of the ion beam position through the perpendicular gas jet while monitoring the beam-loss rate. The target area density of 10$^{14}$\,particles/cm$^2$ was chosen such that charge-changing processes occuring at the intersection point of the gas jet and the ion beam were clearly the dominant beam-loss contribution compared to interactions with the residual gas along the beam line and recombination in the electron cooler. In the energy range of interest, the most relevant charge exchange processes occurring in ion-atom collisions at the gas target are REC and NRC of a target electron. While for low energies NRC dominates the total capture cross section, it is overtaken by REC with increasing collision energies. By passing through a bending magnet downstream to the gas target, down-charged projectile ions were separated from the reference charge state and subsequently stopped by a multi-wire proportional counter (MWPC)~\cite{KLEPPER2003}. Projectiles decelerated to energies below roughly 10\,MeV/u are no longer able to penetrate through the 25\,$\mu$m stainless steel foil separating the ESR vacuum from the MWPC detector housing. Therefore, the measurement of the electron-capture cross-section had to rely on an indirect technique using an array of three high-purity germanium (HPGe) X-ray detectors that were positioned around the interaction zone of the ion beam and the gas target. Each of these detectors covered a solid angle of about 10$^{-2}$\,sr and they were placed with respect to the ion beam axis at 35$^\circ$, 60$^\circ$ and at 90$^\circ$, see figure~\ref{fig1} for details.

\begin{figure}[tp]
\begin{center}
\includegraphics[width=8.6cm]{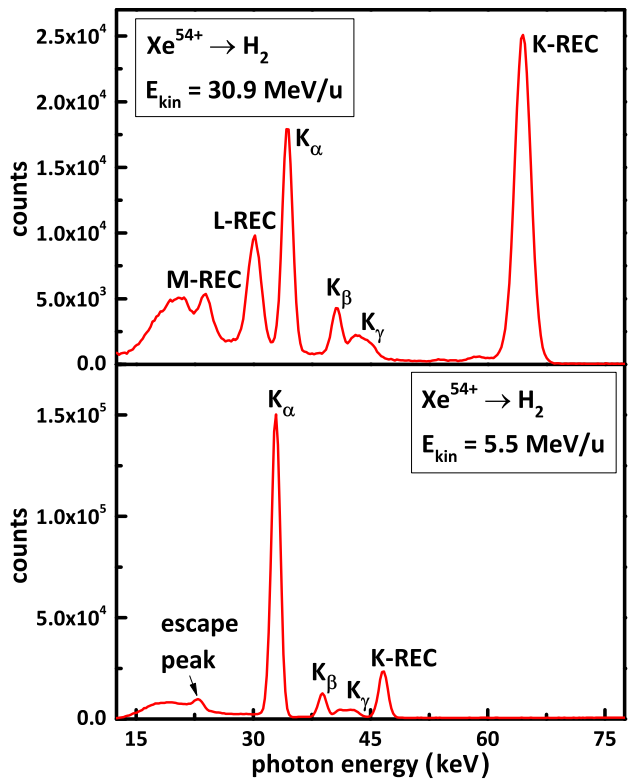}
\end{center}
\caption{X-ray spectra resulting from the collision of bare xenon ions with a H$_2$ gas target, recorded by HPGe X-ray detectors. The spectra were obtained at an observation angle of 60$^\circ$ at the highest and the lowest ion beam energies used in this study. Note that in the laboratory system the peak positions are Doppler shifted with respect to the transition energies in the emitter system. 
\label{fig2}}
\end{figure}

As an example of the results, in figure~\ref{fig2}, two X-ray spectra taken at collision energies of 30.9 MeV/u (top) and 5.5 MeV/u (bottom) at an observation angle of 60$^\circ$ are shown.
All relevant features in these spectra can be attributed to either the REC process (i.e.\ K-REC denotes the radiation emitted by REC into the projectile K~shell, L-REC refers to capture into the L~shell and so on) or characteristic K transitions into the ground state (K$_\upalpha$, K$_\upbeta$, etc.). The difference in the intensity ratios of the REC and the K radiation for both collision energies results from NRC dominating the population of excited projectile states at low collision energies, whereas at the higher energy the REC process prevails. Also visible is an escape peak resulting from the germanium K radiation leaving the detector crystal. All spectra were corrected according to the energy-dependent detection efficiency~$\varepsilon$ of the respective HPGe detector. The efficiency correction was obtained by modeling the detector response based on the EGS5 Monte Carlo code for photon and electron transport in matter~\cite{egs5}. This code is able to reproduce all relevant features of the detector response as previously demonstrated in \cite{Weber_2011}.

For a hydrogen-like high-Z system multi-photon transition rates are relatively small compared to single-photon transitions, the most important being the two-photon 2E1 transition from the $ 2\text{s}_{1/2} $ to the $ 1\text{s}_{1/2} $ state, which has for xenon a rate of about $ 1.9 \times 10^{11} \, \text{s}^{-1} $, while the single-photon M1 transition rate between the same states is about $ 6.3 \times 10^{11} \, \text{s}^{-1}$ \cite{volotka2019}. Furthermore, in a one-electron system Auger decay of excited states is not possible. As a consequence, for almost all electrons that are captured into excited projectile states the transition to the ground state is accompanied by emission of single-photon K radiation. Therefore, the K$_{\upalpha,\upbeta,\upgamma,...}$ emission cross section integrated over the complete solid angle is a measure of the electron-capture cross-section summed over all principal quantum numbers $n$, with $n>1$. A peculiarity of the two spectra in figure~\ref{fig2} lies in the increase of the K radiation intensity relative to the REC peaks as the ion beam energy is decreasing. This illustrates the fact that at low beam energies the NRC process is the dominating capture process. Note that in the REC process electrons are most likely captured directly into low-$n$ states, in particular into the K shell \cite{Eichler2007}. In contrast, in the NRC process electrons are captured mainly into those projectile states whose momentum distribution has a high overlap with the initial momentum distribution, that is given by the target electrons' intrinsic momentum distribution convoluted and shifted by the relative momentum in the collision \cite{Ichihara1994}. For the present collision parameters this leads predominantly to capture into high-$n$ states, which can be seen in Fig. 2, which subsequently undergo cascades of decays towards the ground state, resulting in intense emission of K radiation. 

By normalizing the intensity $I_{\textrm{K}}$ of the peaks formed by the K radiation to the intensity of the K-REC radiation $I_{\textrm{REC}}$ the K emission cross section $\sigma_{\textrm{K}}$ is expressed in terms of the angular differential cross section $\sfrac{\textrm{d}\sigma_{\textrm{\tiny{REC}}}}{\textrm{d}\Omega}$ of the REC process as follows:
\begin{equation}
\label{cs_K}
\frac{\textrm{d}\sigma_\textrm{\tiny{K}}}{\textrm{d}\Omega}=\frac{I_{\textrm{\tiny{K}}}}{I_{\textrm{\tiny{REC}}}}\frac{\varepsilon_{\textrm{\tiny{REC}}}}{\varepsilon_{\textrm{\tiny{K}}}}\frac{\textrm{d}\sigma_{\textrm{\tiny{REC}}}}{\textrm{d}\Omega}\,,
\end{equation}
with $\varepsilon_{\textrm{\tiny{K}}}$ and $\varepsilon_{\textrm{\tiny{REC}}}$ being the detector efficiency at the energy of the K radiation and K-REC radiation, respectively.

A similar normalization procedure to obtain emission cross sections of spectral features relative to the REC peaks has already been used for example in~\cite{Ludziejewski98,Herdrich2017}. It results in a cancellation of sizeable sources of uncertainty of the experimental setup such as the solid angle covered by each detector. The necessary K-REC cross section data, differential in emission angle and photon energy, was produced using the \texttt{RECAL} program~\cite{Weber2015,Herdrich2017}. This algorithm performs interpolations between precalculated radiative recombination (RR) differential cross section values, produced by a code provided by A. Surzhykov et al. \cite{Surzhykov2001}, and convolutes the resulting data with the tabulated Compton profile of the target atom~\cite{BIGGS1975}. The reduction of the REC process to the RR process, which in turn can conveniently be described as the time-inverse process of photoionization, is justified as long as the target electrons' initial binding energy and momentum distribution is negligible compared to the energy and momentum transfer during the capture process, i.e. the initial electron can be treated as quasi-free. This requirement is fulfilled in the present case of a strongly asymmetric collision system, thus enabling the approximation of the REC process as the time-inverse of the well-understood photoionization process, see \cite{Eichler2007} for details. In the aforementioned code the photoionization cross section is obtained within the framework of the Dirac theory for bound and free states, while assuming a point-like nucleus, and by making use of the partial-wave expansion of the continuum electron wave-function. The coupling to the electromagnetic radiation is treated within the framework of the first-order perturbation theory. In order to compute all the angular differential properties, moreover, the density matrix theory is applied. Based on this approach the K-REC differential cross section is expected to be predicted with an accuracy of a few percent \cite{surzhykov}. On the other hand, basically all measurements of the absolute REC cross section reported in the literature exhibit uncertainties in the range of 20\,\% to 50\,\%~\cite{Stoehlker1995,Eichler2007}, thus limiting the experimental verification of any theoretical treatment to this level. In general, an accurate experimental determination of absolute cross section values is challenging since target densities, absolute beam intensities and beam-target overlap have to be determined precisely and also a detailed knowledge of the outgoing particle detection efficiency is required. 

In the emitter system the angular differential cross section of an electric dipole transition is linked to the total transition cross section from the initial to the final state $ \sigma_{\textrm{E1}}^{\textrm{i}\rightarrow\textrm{f}} $ by \cite{Surzhykov2006}:

\begin{equation}
\label{alignment}
\frac{\textrm{d}\sigma_\textrm{\tiny{E1}}}{\textrm{d}\Omega}=\frac{\sigma_{\textrm{E1}}^{\textrm{i}\rightarrow\textrm{f}}}{4\pi} \left (1 + \beta^{\textrm{\tiny{eff}}}A\left(1-\sfrac{3}{2}\sin^2\theta \right)\right)\,,
\end{equation}

where $\beta^{\textrm{\tiny{eff}}}$ is the so-called effective anisotropy parameter that is non-zero for transitions from initial states with angular quantum numbers $j\!>\!\sfrac{1}{2}$ and $A$ is the alignment parameter that takes a non-zero value if the initial state exhibits a non-statistical population of the magnetic substates. Note that a non-statistical population is a common feature of excited states of highly charged ions produced in collision processes, such as REC \cite{Stoehlker1997} and NRC \cite{Hansen1990}. We adjusted equation~(\ref{alignment}) to the experimental $\sfrac{\textrm{d}\sigma_\textrm{\tiny{K}}}{\textrm{d}\Omega}$ data points by treating the total cross section and the product $\beta^{\textrm{\tiny{eff}}}A$ as free parameters, while taking into account the relativistic transformation of the observation angle and the solid state element from the laboratory to the emission system. As mentioned above, the overwhelming majority of electrons captured to excited states of the projectile will decay to the ground state with the emission of K radiation. Thus, the obtained K emission cross section is in good approximation equal to the total electron-capture cross-section for all projectile states with $n>1$. However, for some energies only spectral data from the detectors at $60^\circ$ and $90^\circ$  was available. These observation angles are too close to each other to extract the underlying angular distribution in a  meaningful way. Nevertheless the angular distribution where all three detector positions are available exhibits only a small degree of anisotropy. In fact, all obtained anisotropy parameters are within 1\,$\sigma$ compatible with zero. This is also expected from first principles as the anisotropic $ 2\text{p}_{3/2}  \rightarrow 1\text{s}_{1/2}$ transition is superimposed by the isotropic $ 2\text{p}_{1/2}, 2\text{s}_{1/2}  \rightarrow 1\text{s}_{1/2}$ transitions and also the alignment of the $2\text{p}_{3/2}$ state by direct population is subsequently diluted by cascade feeding. Thus, where the experimental data was limited to two observation angles we approximated the total cross section by averaging the angular-dependent emission cross sections measured by both detectors, implicitly assuming an isotropic emission pattern. A conservative estimate of the overall uncertainty including systematics (the detector efficiency) of the obtained total emission cross-section for all collision energies amounts to $\pm 10$\,\%.

\section{Theoretical approaches for electron capture}

In the following, various theoretical methods available for cross section calculations for the processes of radiative and non-radiative electron capture from low-$Z$ targets by heavy projectiles are briefly described. In addition the Schlachter formula, which yields an empirical estimate of the total cross section for electron capture based on a set of experimental data that was available in the early 1980s, is presented. All these approaches are then compared to the experimental results. It should be noted that a variety of other methods for the treatment of the non-radiative electron capture process exists, e.g. $n$-particle Classical Trajectory Monte Carlo (nCTMC) simulations, Continuum Distorted Wave (CDW) theory, coupled channel method, just to name a few. However, in this work we take into account only those treatments that are readily available and frequently used within the community for prompt and pragmatic estimates of charge-exchange rates and ion beam lifetimes.\\

\paragraph*{Radiative electron capture}
The cross section for electron capture, due to the REC process integrated over all photon emission angles, was obtained by two separate methods. To obtain the cross sections for capture into projectile shells with principle quantum numbers $n=2,3$, an interpolation was performed on an extensive tabulation of cross section values published by Ichihara and Eichler~\cite{rec}. These values are based on fully relativistic calculations which also include the effects of the finite nuclear size and all multipole orders of the photon field for RR into the K, L and M shell of bare ions. The only difference between this approach and the one on which the RECAL code is based, is the consideration of the finite nuclear size in the former. Therefore, the cross sections from Ichihara and Eichler ~\cite{rec} can be considered to be more complete in principle. However, for the present level of accuracy the difference is not significant. Moreover, for the shells $n>3$ a non-relativistic approach was used which is generally applicable when both the collision energy and the binding energy of the captured electron are considerably smaller than the electron rest mass. This treatment of the RR process is based on recurrence relations as described in \cite{Eichler2007}.\\

\paragraph*{Non-radiative electron capture according to the eikonal approximation}
The Eikonal approximation is known to describe total electron-capture cross-sections for asymmetric collision systems at high energies within a factor two to three. Within the Eikonal approximation, one center is treated in first order $Z \alpha$ whereas the other center is described non-perturbatively. A closed formula for the relativistic eikonal approximation of electron capture from the 1s state of a hydrogen-like target into the 1s shell of an initially bare projectile was derived by Eichler \cite{Eichler85}. It has the form
\begin{equation}
\label{eikonal}
\begin{aligned}
\sigma_{1\textrm{s},1\textrm{s}}^{\textrm{eik}}& ={a_{0}}^2\pi\frac{2^{8}\left(Z_\textrm{P} Z_\textrm{T}\right)^5}{5v^2\left({Z_\textrm{T}}^2+{p_{-}}^2\right)^5}\frac{1+\gamma}{2\gamma^2}\frac{\pi{\eta}{Z^{\prime}_\textrm{T}}}{\sinh\left({\pi{\eta}{Z^{\prime}_\textrm{T}}}\right)}\\
& \times e^{-2{\eta}{Z^{\prime}_\textrm{T}}\tan^{-1}\left(-p_{-}/Z_\textrm{T}\right)}\left(S_\textrm{eik}+S_\textrm{magn}+S_\textrm{orb}\right)\,,
\end{aligned}
\end{equation}
where $Z_\textrm{P}$ and $Z_\textrm{T}$ denote the atomic numbers of the projectile, respectively target, $v$ is the collision velocity and $\gamma$ is the associated relativistic Lorentz factor, while $a_{0}$ denotes the Bohr radius. For a detailed explanation of the other parameters in equation~(\ref{eikonal}) the reader is referred to the original publication \cite{Eichler85}. The parameter $Z^{\prime}_\textrm{T}$, which in the present work was chosen as $Z^{\prime}_\textrm{T}=Z_\textrm{T}$, represents the potential of the target system in a final-state interaction with the captured electron now being bound to the projectile. For the present case, where the projectile potential is the stronger one, the `post' version of the eikonal approximation was adopted by exchanging $Z_\textrm{P} \leftrightarrow Z_\textrm{T}$ and substituting $Z^{\prime}_\textrm{T}$ with $Z^{\prime}_\textrm{P}$. It was shown by Meyerhof et al.~\cite{Meyerhof85} that the eikonal cross section averaged over a complete principal shell scales with $\sfrac{Z}{n}$ for initial and final states, thus enabling the extension of equation~(\ref{eikonal}) to capture from and into shells having arbitrary quantum numbers $n$ by making the substitution $Z_\textrm{T} \rightarrow \sfrac{Z_\textrm{\tiny{T}}}{n_\textrm{\tiny{T}}}$, respectively $Z_\textrm{P} \rightarrow \sfrac{Z_\textrm{\tiny{P}}}{n_\textrm{\tiny{P}}}$. For practical purposes a cut-off value is necessary for the highest projectile shells to be considered, which in this work was set to $n_{\textrm{cut}}=50$. It should be noted that the underlying approximations for the closed formula presented by Eichler are only valid for collision velocities higher than the orbital velocities of the initial and the final state of the electron, which is for most collision energies under investigation clearly not the case with respect to the projectile K and L shell. However, as the total NRC cross section is dominated by capture into those shells that have the largest overlap with the initial electron momentum distribution (taking into account the collision velocity), the contribution by capture into the K and L shell is rather small at the relatively low collision energies which are of interest in this work.\\

\paragraph*{Non-radiative electron capture according to the \texttt{CAPTURE} code}
This code is based on the normalized Brinkman-Kramers (NBK) approximation in the impact parameter representation \cite{Shevelko2001,Shevelko2004}. Like the eikonal approximation it utilizes hydrogen-like radial wave functions to describe the initial and the final state of the captured electron. The total NRC cross section is given as a sum of partial cross sections $\sigma_{n^{\textrm{f}},n^{\textrm{i}}}$ for capture from all occupied target electron shells with the principle quantum number $n^{\textrm{i}}$ into all possible final projectile states having the principal quantum number $n^{\textrm{f}}$ as follows:

\begin{equation}
\label{capture}
\begin{aligned}
\sigma^{\texttt{CAPTURE}} & =\sum_{n^{\textrm{f}}=1}^{n^{\textrm{f}}_{\textrm{cut}}}{\sum_{n^{\textrm{i}}=1}^{n^{\textrm{i}}_{\textrm{max}}}{\sigma_{n^{\textrm{f}},n^{\textrm{i}}}}}\,, \\
\sigma_{n^{\textrm{f}},n^{\textrm{i}}} & = 2\pi\int_{0}^{b_{\textrm{max}}}P^{\textrm{norm}}_{n^{\textrm{f}},n^{\textrm{i}}}\left(b\right)b\,db\,, \\
P^{\textrm{norm}}_{n^{\textrm{f}},n^{\textrm{i}}}\left(b\right) & = \frac{P_{n^{\textrm{f}},n^{\textrm{i}}}\left(b\right)}{1+\sum_{n^{\prime} \neq n^{\textrm{f}}} P_{n^{\prime},n^{\textrm{i}}}\left(b\right)}\,,
\end{aligned}
\end{equation}

where $P_{n^{\textrm{f}},n^{\textrm{i}}}\left(b\right)$ is the probability according to the Brinkman-Kramers (BK) treatment for capture of an electron from the (initial) target shell $n^{\textrm{i}}$ into the (final) projectile shell $n^{\textrm{f}}$, depending on the impact parameter~$b$. The main feature of this approach is that the normalized capture probability $P^{\textrm{norm}}$ is always less than unity, making it possible to use the NBK approximation even at lower energies which are not accessible with the pure BK approximation.

\begin{figure}[tb]
\begin{center}
\includegraphics[width=8.6cm]{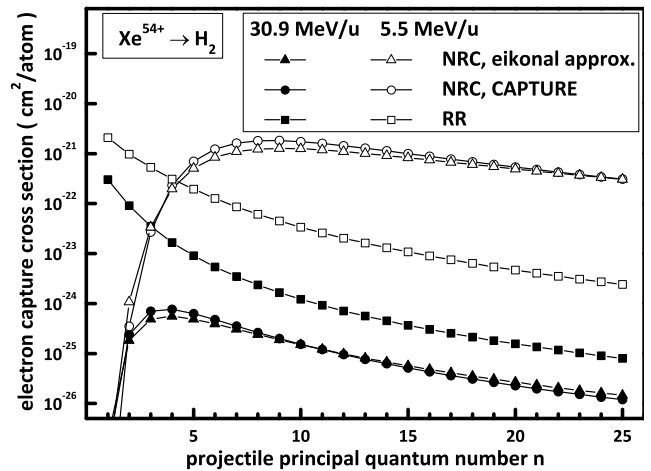}
\end{center}
\caption{Theoretical electron-capture cross-sections against the projectile principle quantum number $n$ for the collision of Xe$^{54+}$ with a H$_{2}$ target at kinetic beam energies of 5.5 and 30.9 MeV/u. The cross sections are given per target molecule constituent, i.e. per atom. The RR cross section data was produced with a non-relativistic treatment based on recurrence relations, while the NRC cross section values were calculated using the eikonal approximation as well as the \texttt{CAPTURE} code. It is found that both NRC treatments yield very similar results, with the \texttt{CAPTURE} code predicting slightly larger cross section values than the eikonal approximation.
\label{fig3}}
\end{figure}

In figure~\ref{fig3} electron-capture cross-sections are presented which result from the aforementioned treatments for the collision of bare xenon projectiles with hydrogen at the lowest and at the highest collision energy under investigation in this work. As can be seen both NRC treatments predict very similar cross sections, with the \texttt{CAPTURE} code predicting slightly larger values than the eikonal approximation. In fact, it is known that setting $Z^{\prime} = 0$ in the eikonal approximation results in a capture cross section identical to the NBK value \cite{Eichler85}. While the RR/REC process is dominated by the capture into low-$n$ states of the projectile, the NRC process exhibits the largest cross section for capture into those shells having the largest overlap with the initial electron momentum distribution. Since hydrogen has a very narrow intrinsic momentum distribution, the initial electron momentum is effectively defined by the collision velocity. As a consequence, the NRC cross section peaks at principal quantum numbers of the projectile that roughly correspond to orbital velocities around half the collision velocity. Thus, for probing the predictive power of NRC treatments in the context of the present work, it is reasonable to focus on capture into excited states and to neglect the contribution of the K shell. The theoretical data integrated over electron capture into all projectile principle quantum numbers $n\geq2$ is presented in the upper picture of figure~\ref{fig4}, as described in section \ref{res}.\\

\paragraph*{Total electron capture according to the Schlachter formula}
An empirical cross section formula for single electron capture was obtained by Schlachter et al.~\cite{Schlachter83} based on a large data set covering collision energies between 0.3\,MeV/u and 8.5\,MeV/u and initial projectile charge states up to $59+$. This so-called Schlachter formula has the following form:

\begin{equation}
\label{schlachter}
\begin{aligned}
\sigma^{\textrm{Schlachter}} & =\frac{q^{0.5}}{Z_{\textrm{T}}^{1.8}}\frac{1.1\times 10^{-8}}{\tilde{E}^{4.8}}\left(1-e^{-0.037\tilde{E}^{2.2}}\right)\\
& \times \left(1-e^{-2.44\times 10^{-5}\tilde{E}^{2.6}}\right)\left[\sfrac{\textrm{cm}^2}{\textrm{atom}}\right]\,,
\end{aligned}
\end{equation}

where $q$ denotes the projectile charge and the reduced energy $\tilde{E}=E/\left(Z_{\textrm{T}}^{1.25}q^{0.7}\right)$ is derived from the kinetic projectile energy $E$ expressed in units of keV/u. The range of validity is stated as $q \geq 3$ and $10 \leq \tilde{E} < 1000$. The underlying data does not contain highly charged, heavy projectiles with open K and L shells at collision velocities, where the REC process significantly contributes to the total capture cross section. This is the reason why, even though the conditions for $q$ and $\tilde{E}$ are fulfilled (with the exception of the 30.9\,MeV/u data point which corresponds to $\tilde{E}=1900$) the applicability of the Schlachter formula is questionable for the collision system addressed in this work. However, as the Schlachter formula is widely used for pragmatic estimations of the capture cross section in ion-atom collisions, it is important to test its predictive power in a variety of scenarios including also edge cases like the mentioned ones, see the lower picture of figure~\ref{fig4}.

\section{Results and discussion}
\label{res}
\begin{figure}[tb]
\begin{center}
\includegraphics[width=8.6cm]{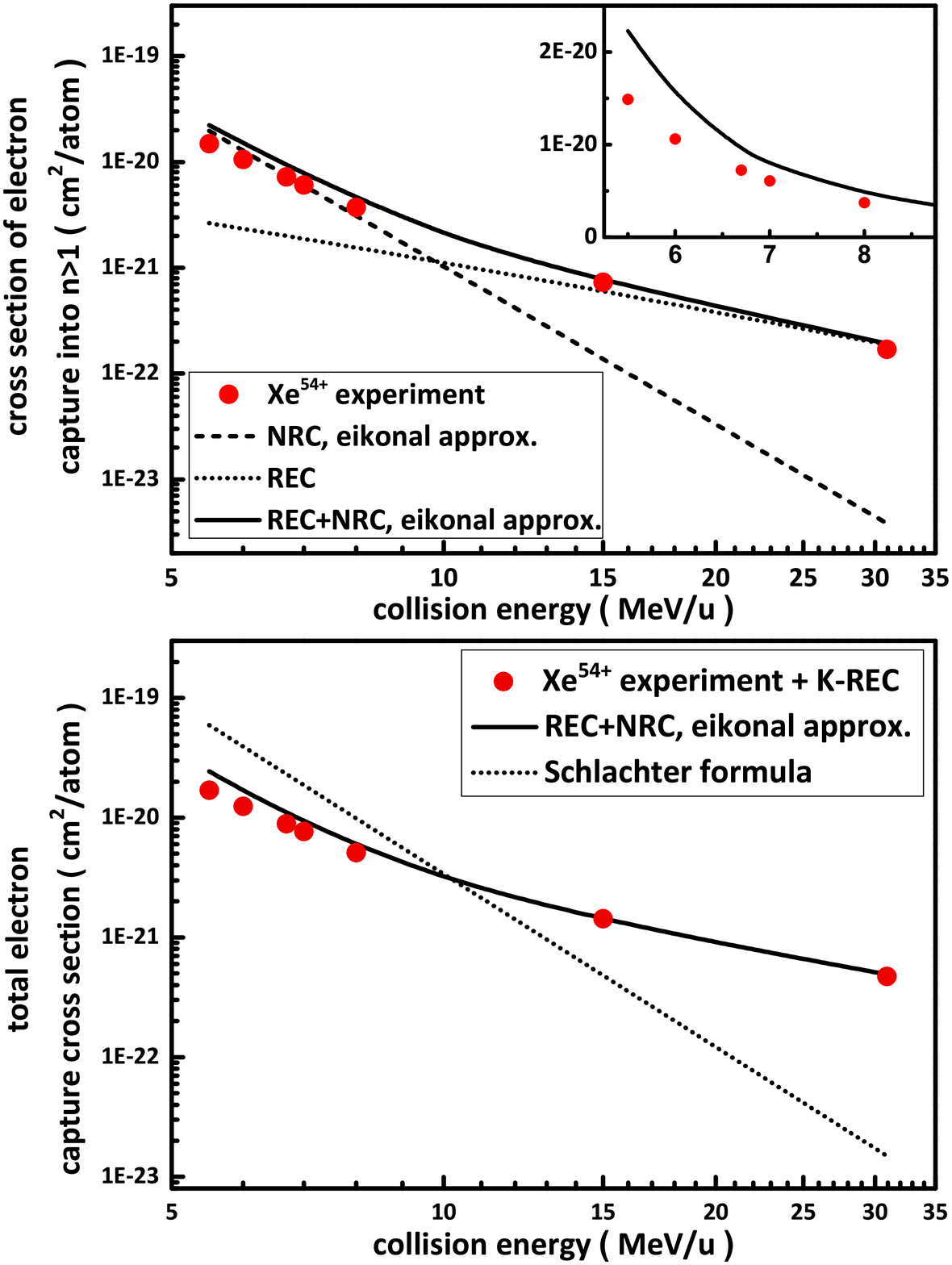}
\end{center}
\caption{Upper picture: cross section of electron capture from hydrogen into excited states of xenon projectiles as a function of the collision energy. The experimental data is shown together with theoretical cross sections for the REC and the NRC process. The inset plot has a linear scaling and shows the systematic overestimation at lower beam energies where the NRC process dominates in more detail. Lower picture: total electron capture cross section, produced by amending both the experimental and theoretical data for capture into excited projectile states with the theoretical K-REC cross section. As can be seen, the Schlachter formula is unable to reproduce the experimental data both at low and at high collision energies.
\label{fig4}}
\end{figure}

The experimental cross section values obtained in this work for electron capture into excited projectile states are presented in the upper picture of figure~\ref{fig4}.
For the 15~MeV/u and 30.9~MeV/u data points, where the REC contribution to the total capture cross section is dominant, a correction for the unobserved capture events due to two-photon decay from the $2\text{s}_{1/2}$ state was applied. Taking into account the direct population by the REC process as well as cascade feeding from all other states with $n=$ 2 and 3 results in an enhancement of the electron-capture cross-section by 3.4\,\% and 5.5\,\%, respectively. For the lower beam energies, a correction is not necessary as the NRC process, in contrast to REC, is only sparsely populating the $2\text{s}_{1/2}$ state. Also shown in the upper picture of figure~\ref{fig4} are theoretical cross sections for the REC and the NRC process. For the latter only the eikonal approximation is shown as the \texttt{CAPTURE} code yields cross section values that are very similar, as can be seen figure~\ref{fig3}, so that both data sets would be hardly distinguishable. The depicted cross section values are also listed in table~\ref{table1}. For the energy region that is not completely dominated by the REC process, it is found that both NRC treatments lead to a systematic overestimation of the experimental data while reproducing the overall shape of the cross section fairly well. As the eikonal approximation yields slightly smaller values it is in marginally better agreement with the experimental results. Nevertheless, a deviation from the experimental findings by 25\% to 50\% is observed, with an increasing trend at lower beam energies. This feature is shown in more detail in the inset plot of the upper picture of figure~\ref{fig4}.

As far as the operation of storage rings like CRYRING@ESR is concerned, the most relevant information on electron capture is the total cross section integrated over capture into all projectile states. It allows the estimation of important beam parameters, such as beam losses caused by charge-exchange processes, making it a crucial input for the planning of future experiments and the design of experimental setups. In the lower picture of figure~\ref{fig4} total electron-capture cross-sections are presented. These were obtained by adding the theoretical K-REC cross section (interpolated from the tabulated values by Ichihara and Eichler~\cite{rec}) to the experimental data and also to the theoretical capture cross sections into projectile states with $n > 1$. It should be noted that for all collision energies considered in this study NRC into the K-shell is negligible, whereas REC into the K shell is the dominant REC contribution, as can be seen in figure~\ref{fig3}.

\begingroup
\begin{table}[tb]
\caption{Experimental electron-capture cross-section data obtained in this work in comparison with the sum of predictions for the NRC and the REC process, as presented in the upper picture of figure~\ref{fig4}. The stated uncertainties reflect the experimental contribution only and do not account for potential theoretical uncertainties in the prediction of the K-REC cross section that was used for normalization, see text for details. 
\label{table1}}
\begin{ruledtabular}
\begin{tabular}{cccc}
\multirow{3}{*}{Collision system}  & \multirow{2}{*}{\makecell{Collision energy \\ (MeV/u)}} & \multicolumn{2}{c}{\makecell{Electron capture \\ into $n>1$ \\ ($10^3$\,barn/atom)}}   \\
 & & Experiment & NRC+REC  \\
\hline
\multirow{7}{*}{Xe$^{54+}$ $\rightarrow$ H$_2$}
&	\phantom{0}5.5	&	14.9\phantom{0}	$ \pm $	1.49	&	22.3\phantom{0}	\\
&	\phantom{0}6\phantom{.0}	&	10.6\phantom{0}	$ \pm $	1.06	&	15.1\phantom{0}	\\
&	\phantom{0}6.7	&	\phantom{0}7.2\phantom{0}	$ \pm $	0.72	&	\phantom{0}9.4\phantom{0}	\\
&	\phantom{0}7\phantom{.0}	&	\phantom{0}6.1\phantom{0}	$ \pm $	0.61	&	\phantom{0}7.9\phantom{0}	\\
&	\phantom{0}8\phantom{.0}	&	\phantom{0}3.7\phantom{0}	$ \pm $	0.37	&	\phantom{0}4.6\phantom{0}	\\
&	15\phantom{.0}	&	\phantom{0}0.73	$ \pm $	0.07	&	\phantom{0}0.74	\\
&	30.9	&	\phantom{0}0.17	$ \pm $	0.02	&	\phantom{0}0.19	\\
\end{tabular}
\end{ruledtabular}
\end{table}
\endgroup

From a practical point of view the comparison shown in the lower picture of figure~\ref{fig4} is the most relevant to assess the predictive power of the various treatments for electron capture. As can be seen, in the lower energy region where NRC is the dominating capture process, once again the REC+NRC capture cross section is systematically larger by up to 50\% when compared to the experimental values. Nevertheless, when taking into account the experimental uncertainties, there is a reasonable qualitative agreement overall. In contrast, the Schlachter formula fails to reproduce the REC contribution which results in a severe underestimation of the total capture cross section on the high energy part of the data set, while it significantly overestimates the capture cross section at lower energies, where NRC dominates. More specifically, the Schlachter formula exhibits a similar energy-dependence as predicted by the eikonal approximation and the \texttt{CAPTURE} code, but yields absolute values that are roughly by a factor of 3 larger, which is clearly in disagreement with the experimental data. At first glance, the apparent inapplicability of the empirical formula in this study is not surprising since the underlying data set does not contain heavy, highly charged ions with open K and L shells such as bare xenon. This explains why the REC contribution is not reproduced well by the Schlachter formula. However, in contrast to the REC process, these open inner shells of the projectile do not contribute significantly to the total NRC cross section in the energy range under investigation. Thus, it is notable that a large deviation is also found in the NRC-dominated energy region.

In this context it is also worth noting that in a previous electron capture study using decelerated highly charged ions, namely hydrogen-like germanium in collision with a neon target, good agreement with the Schlachter formula was found~\cite{Stohlker_1992}. In that study the experimental data was also well reproduced by an $n$-particle classical trajectory Monte Carlo calculation. In contrast, the eikonal approach resulted in an overestimation of the capture cross section by about a factor of 2. However, one has to keep in mind that germanium ions colliding with neon atoms is a significantly more symmetric collision system than xenon colliding with hydrogen which was studied in the present case. Summarizing, given the range of relevant collision parameters, further experimental studies of cross section data for electron capture by highly charged, medium to high-$Z$ ions at low collision energies are necessary to draw definite conclusions on the reliability and range of applicability of the various theoretical and empirical predictions.

\section{Summary and outlook}
The cross section for electron capture into excited projectile states was measured in collisions of Xe$^{54+}$ with a H$_2$ target in a low-energy regime not accessible up to now. At such low velocities highly-charged heavy ions have charge states much higher than their respective equilibrium charge-state, and as a consequence electron capture from the residual gas constituents is the dominant beam-loss process. Thus, precise knowledge of electron-capture cross-sections is crucial for the correct estimation of charge-exchange rates and ion beam lifetimes in accelerators and storage rings. The obtained cross section values were compared to theoretical treatments of the non-radiative capture and the radiative electron capture, as well as an empirical formula for total electron capture. It is found that non-radiative electron-capture cross-section values predicted by the eikonal approximation and the \texttt{CAPTURE} code are in reasonable qualitative agreement with the experimental findings, even though with decreasing beam energy the electron capture is overestimated by up to 50\%. This still reasonable agreement is quite remarkable considering that both theories applied are high-energy approximations. For the current beam energy regime and even lower energies, it is evident that more adequate low-energy models need to be investigated and applied. Moreover, the commonly used empirical Schlachter formula significantly overestimates the total capture cross section at low collision energies, where non-radiative capture dominates.

\begin{acknowledgments}
This project has received funding from the European Research Council (ERC) under the European Union's Horizon 2020 research and innovation programme (grant agreement No 682841 ``ASTRUm''). Moreover, this work was supported by the Helmholtz International Center for FAIR (HIC for FAIR), by the Bundesministerium f\"ur Bildung und Forschung (BMBF) (05P15RFFAA, 05P15RGFAA), by the Science and Technology Facilities Council (STFC) UK (ST/L005824/1, ST/M001652/1, ST/M006085), by the Helmholtz-CAS Joint Research Group (HCJRG-108), and by the Helmholtz-OCPC Postdoctoral Program 2017 (GSI08).

FMK acknowledges support by the BMBF Verbundprojekt 05P2018 (ErUM-FSP T05).
CLW acknowledges support by the European Research Council (grant agreement ERC-2015-StG Nr. 677497 ``DoRES'').
SD gratefully acknowledges the support provided by the Alexander von Humboldt Foundation and the Jordanian Scientific Research Support Fund under grant \#Bas/2/4/2014.
SYuT acknowledges the support by the DAAD through Mendeleev grant, SPbU (28999675).
YuAL acknowledges the support by the DAAD through Programm des projektbezogen Personenaustausch (PPP) with China [Project ID 57389367].
TSz acknowledges support from Helmholtz Association (ERC-RA-0016).

\end{acknowledgments}

\bibliography{references}

\end{document}